\documentclass[aps,showpacs,twocolumn,preprintnumbers]{revtex4}

\usepackage{amsmath,amssymb} \usepackage{graphicx}
\usepackage{psfrag,color,hyperref}

\renewcommand{\vec}{\mathbf}

\begin{document}

\title{The effect of internal and global modes on the radial distribution function of confined semiflexible polymers}

\author{F. Th\"uroff} \author{F. Wagner} \author{E. Frey}

\affiliation{                    
  Arnold Sommerfeld Center for Theoretical Physics (ASC) and Center for NanoScience (CeNS), LMU M\"unchen, Theresienstra{\ss}e 37, 80333 M\"unchen, Germany\\
}

\begin{abstract}

The constraints imposed by nano- and microscale confinement on the conformational degrees of freedom of thermally fluctuating biopolymers are utilized in contemporary nano-devices to specifically elongate and manipulate single chains. A thorough theoretical understanding and quantification of the statistical conformations of confined polymer chains is thus a central concern in polymer physics.
We present an analytical calculation of the radial distribution function of harmonically confined semiflexible polymers in the weakly bending limit. Special emphasis has been put on a proper treatment of global modes, \textit{i.e.} the possibility of the chain to perform global movements within the channel.
We show that the effect of these global modes significantly impacts the chain statistics in cases of weak and intermediate confinement. Comparing our analytical model to numerical data from Monte Carlo simulations we find excellent agreement over a broad range of parameters.

\end{abstract}

\pacs{87.15.ad,82.35.Lr,36.20.Ey} \date{\today}

\preprint{LMU-ASC 04/10}

\maketitle

Understanding the statistical properties of biopolymers on the single molecule level has been one major objective of polymer research over the past decades. 
While our theoretical knowledge about semiflexible chain molecules in free space is developed to a fairly sophisticated level, a steadily growing theoretical interest 
in the mechanical properties of semiflexible polymer solutions \cite{ISI:A1996TT68800035,ISI:000076389300040,PhysRevLett.81.2614,ISI:000167623900036,ISI:000249809400005}, 
and networks \cite{PhysRevLett.75.4425,ISI:000185485700046,ISI:000185485700045,PhysRevLett.97.105501,bausch,ISI:000251885000011}, 
as well as the impetus from nano-technological applications \cite{Reisner,ISI:000223000200026} 
calls for an expansion of this knowledge to the case of polymers trapped within some confining environment.
Odijk \cite{odijk} and de Gennes \cite{deGennesIntrotoScaling} laid the groundwork for further analytical treatments of confined chain molecules by providing very successful scaling theories for confined semiflexible (Odijk) and flexible (de Gennes) chains.
Recent advances in photolithographic techniques allowed for the fabrication of channel-like structures down to nanometer scales, thus enabling the experimentalist to investigate the effect of confinement on single chain molecules. Channels of different sizes have been used to directly observe the fluctuations of F-actin and DNA molecules and to investigate the respective statistical properties within Odijk's \cite{ISI:000235147000007,ISI:000255867900010,ISI:000233225600061} and de Gennes' \cite{ISI:000223000200026} scaling regime and at the crossover between both \cite{Reisner}. 

On the theoretical side, questions related to thermodynamic properties and the end-to-end distance statistics of polymers in confining geometries have largely been addressed by means of computer simulations \cite{wagner-2007-75,cifra,yang:011804,TWBurkhardt}.
A recent contribution to analytically capture the radial distribution function (RDF) \cite{levi-2007-78} of semiflexible chains in cylindrical confinements was based on the physical idea that the polymer's transverse fluctuations are effectively suppressed along the contour, if the confining channel is sufficiently narrow. In order to facilitate calculations, this approach then employed \textit{torqued ends} boundary conditions \cite{ISI:000073393500037} within a channel-fixed reference frame\footnote{Actually the RDF turns out to be the same for \textit{torqued} and \textit{hinged ends} boundary conditions, which is why---occasionally---we will simply refer to \textit{hinged ends}.}, assuming that the particular choice of boundary conditions is of minor importance for the determination of the chain's end-to-end distance statistics.
While feasible in a strong confinement regime, where the polymer almost perfectly aligns with the symmetry axis of the channel, the chain statistics predicted by this approach is at variance with our results from Monte Carlo simulations for intermediate and weak confinement strengths.

In this letter we go beyond this strong confinement approximation and provide a thorough analysis of the end-to-end distance statistics of semiflexible chains, trapped in cylindrically symmetric confinements, based on analytical and numerical methods. Special attention is paid to the role of global chain movements, which is technically captured by means of a careful choice of boundary conditions. As a central result we will present a refined version of the RDF which is shown to reproduce the chain statistics even in cases of weak and intermediate confinement. In contrast to the RDF obtained in ref. \cite{levi-2007-78}, our approach is capable of reproducing the correct form of the RDF in the particular limit of zero confinement \cite{wilhelm-1996-77,PhysRevE.74.041803}. In order to test our analytical predictions we employed a standard Monte Carlo (MC) scheme \cite{wagner-2007-75}. For this purpose we used a discretized version of the wormlike chain model, consisting of $N$ inextensible segments of length $N^{-1}$ (where the contour length $L=1$ is set to unity). An energetic penalty, proportional to the squared transverse displacement from the symmetry axis of the cylindrical potential, was calculated at the ends of each segment to mimic the presence of confining channel walls.
Self-avoidance effects, which are not important for sufficiently stiff chains, have been neglected.
During the simulation process the chain's ends were assumed to be completely free.

To reveal the importance of boundary effects in the context of end-to-end distance statistics, we start with a short review of some qualitative features concerning the connection between the chain's shape and the constraints at its boundaries. Reference \cite{wagner-2007-75} gives a classification of the shapes, semiflexible polymers in intermediate confinements actually attain. There it was shown that the distribution of transverse displacements for chains whose end-monomers are free to move within the channel resembles a bone-like shape with particularly enhanced fluctuations within a boundary layer of size $\sim L_d$ ($L_d$: Odijk's deflection length \cite{odijk}). In contrast, \textit{hinged ends} boundary conditions give rise to qualitatively different, cigar-like shapes of the polymer. This, as will be detailed in the course of our quantitative discussions, results in significantly larger end-to-end distances in the intermediate confinement regime and significantly smaller end-to-end distances in the limit of vanishing confinement respectively. In the limit of strong confinement the chain's shape approaches a straight line in both cases and the quantitative differences related to the end-to-end distance statistics vanish.

\begin{figure}
	\centering
		\includegraphics[width=.45\textwidth]{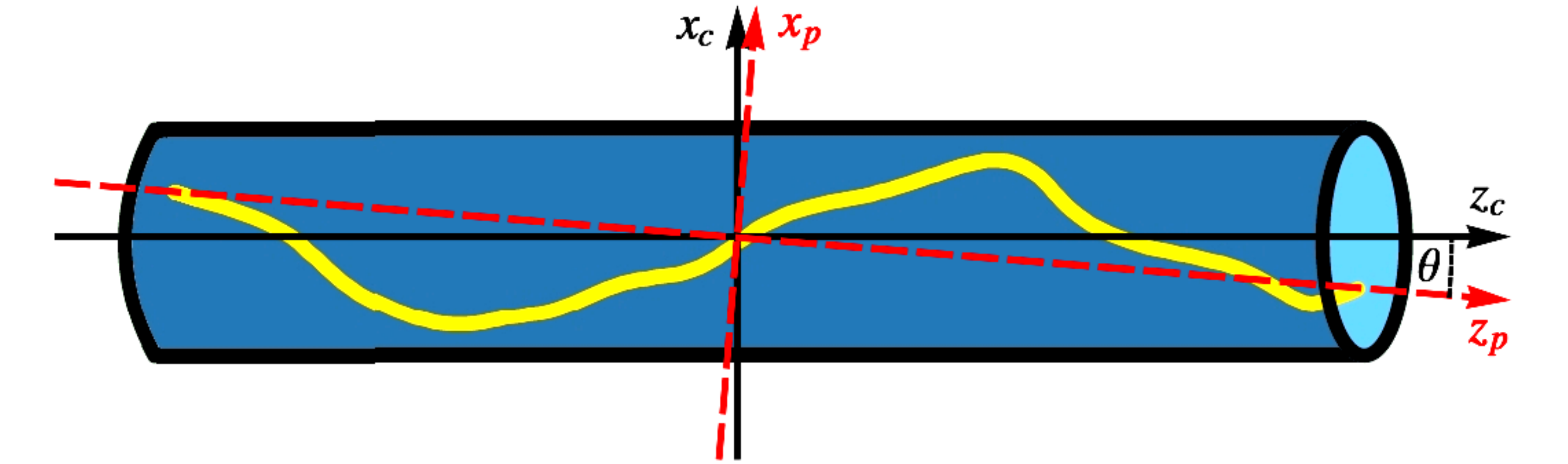}
	\caption{Definition of polymer- and channel fixed reference frame $\Sigma_p$ (red, dashed) and $\Sigma_c$ (black, solid). $\theta$ measures the angle between the two $z$-axes.}
	\label{fig:intro}
\end{figure}

The RDF of the system may be determined by means of an integral over all configurations $\vec{r}(s)$ consistent with one particular value $r$ of the end-to-end distance $R$, and weighted by the appropriate Boltzmann factor:
\begin{equation}
\label{eq:PI}
\mathcal{G}(r)\propto\int\mathcal{D}[\vec{r}(s)]\delta\left(R[\vec{r}(s)]-r\right)\exp\left(-\mathcal{H}[\vec{r}(s)]\right).
\end{equation}
Here and in the following all energies are measured in units of $k_BT$ and all lengths are measured in units of the contour length $L$. The system's Hamiltonian $\mathcal{H}=\mathcal{H}_{\text{bnd}}+\mathcal{H}_{\text{pot}}$ consists of two parts. The bending part is given by a standard wormlike chain Hamiltonian \cite{KratkyPorod} in its continuous form \cite{saito}
\begin{equation}
\label{eq:Hbnd}
\mathcal{H}_{\text{bnd}}=\frac{1}{2\epsilon}\int_{-1/2}^{1/2}ds\,\left[\partial^2_s\vec{r}(s)\right]^2,
\end{equation}
where $\epsilon$, the inverse persistence length, measures the chain's flexibility. The second part accounts for the channel confinement, which is modeled by a harmonic potential
\begin{equation}
\label{eq:Hpot}
\mathcal{H}_{\text{pot}}=\frac{c^4}{2\epsilon}\int_{-1/2}^{1/2}ds\,\left[\vec{r}_{\perp}(s)\right]^2.
\end{equation}
Here $\vec{r}_{\perp}$ denotes the transverse displacements from the symmetry axis of the channel and the collision parameter $c=L_d^{-1}$ may be pictured as the average number of collisions of the contour with the channel walls.

In order to proceed analytically, we investigate eq. \eqref{eq:PI} in Laplace space, \textit{i.e.} we calculate a generating function
\begin{equation}
\label{eq:PILT}
\mathcal{L}(\xi)\equiv\mathfrak{L}\left[\mathcal{G}(r)\right]\propto\int\mathcal{D}[\vec{r}(s)]\exp\left(-\mathcal{E}[\vec{r}(s),\xi]\right),
\end{equation}
where $\mathfrak{L}\left[.\right]\equiv\int_0^{\infty}d\rho e^{-\xi\rho}(.)$ denotes the Laplace transform with respect to the stored length $\rho\equiv1-r$. Thence the RDF itself is given by
\begin{equation}
\label{eq:invLT}
\mathcal{G}(r)=\mathfrak{L}^{-1}\left[\mathcal{L}(\xi)\right]=\int_{-i\infty}^{i\infty}\frac{d\xi}{2\pi i}e^{\xi\rho}\mathcal{L}(\xi).
\end{equation}
In the limit of strong confinements, where the polymer's transverse fluctuations from the channel axis become sufficiently small ($|\vec{r'}_{\perp}|^2=\mathcal{O}(\epsilon/c)\ll1$), one may track the polymer's position with respect to the channel axis and employ a weakly bending rod (WBR) approximation \cite{wilhelm-1996-77}. Then, the exponential $\mathcal{E}$ attains the following form
\begin{equation}
\label{eq:exp}
\begin{split}
\mathcal{E}[\vec{r}(s),\xi]=&\frac{1}{2\epsilon}\left\{\vec{r}_{\perp}''\vec{r}_{\perp}'\bigr|_{-\frac{1}{2}}^{\frac{1}{2}}-\vec{r}_{\perp}'''\vec{r}_{\perp}\bigr|_{-\frac{1}{2}}^{\frac{1}{2}}\right\}+\frac{\xi}{2}\vec{r}_{\perp}'\vec{r}_{\perp}\bigr|_{-\frac{1}{2}}^{\frac{1}{2}}+\\
&+\frac{1}{2}\Bigl\langle\vec{r}_{\perp}\Bigl|\epsilon^{-1}\partial^4_s-\xi\partial^2_s+\frac{c^4}{\epsilon}\Bigr|\vec{r}_{\perp}\Bigr\rangle,
\end{split}
\end{equation}
where $\langle\vec{x}|\vec{y}\rangle\equiv\int_{-1/2}^{1/2}ds\,\left[\vec{x}^t(s)\vec{y}(s)\right]$ abbreviates the $L_2$ scalar product (``$\vec{x}^t$'' meaning ``transpose of $\vec{x}$'') and where primes indicate derivatives with respect to $s$. Note, in particular, that within the WBR approximation the exponential $\mathcal{E}$ becomes a functional of the transverse displacements $\vec{r}_{\perp}(s)$ only. As shortly discussed above, the particular choice of boundary conditions is of minor relevance in that limit. One may, therefore, make the simplest choice of either \textit{hinged} or \textit{torqued ends} boundary conditions in order to make the surface terms in the first line of eq. \eqref{eq:exp} vanish. This then allows to use simple trigonometric eigenfunctions which diagonalize the operator occurring in the second line. Both choices finally lead to the strong confinement approximation of the RDF, determined in ref. \cite{levi-2007-78}.

It must be stressed, however, that the particular form of the exponential $\mathcal{E}$ stated in eq. \eqref{eq:exp} certainly brakes down in the limit of vanishing confinement, where the transverse fluctuations $\vec{r}_{\perp}$ from the channel axis may become large and the WBR approximation fails. Moreover, as detailed above, boundary effects become important outside the strong confinement regime, whence the somewhat artificial choice of \textit{hinged} or \textit{torqued ends} boundary conditions within a channel-fixed reference frame is no longer appropriate. Both problems can be remedied by a proper choice of coordinates. To this end, in addition to the channel-fixed reference frame $\Sigma_c$ used so far (whose $z$-axis coincides with the symmetry axis of the system), we define a second ``polymer-fixed'' reference frame $\Sigma_p$, whose $z_p$-axis shall be defined to connect both end-monomers of the polymer (cf. fig. \ref{fig:intro})\footnote{Note that the axes perpendicular to the $z$-axes of both reference frames may be chosen arbitrarily, due to the rotational symmetry of the problem.}. Within this newly defined frame we may apply Monge parameterization
\begin{equation}\label{eq:Monge}
\mathbf{r}_p(s)=\left(
\begin{array}{c}
\mathbf{r}_{\perp,p}(s)\\
s+\mathcal{O}(\alpha)
\end{array}
\right),\quad s\in\left[-\frac{1}{2},\frac{1}{2}\right],
\end{equation}
where $\alpha=\epsilon/c$ for intermediate (and strong) confinement ($c\gtrsim 1$) and $\alpha=\epsilon$ for vanishing confinement ($c\rightarrow0$). Note that the physically correct boundary conditions within $\Sigma_p$ read
\begin{subequations}
\label{eq:BC}
\begin{eqnarray}
\label{eq:BC1}\vec{r}_{\perp,p}(s)\bigr|_{s=\pm1/2} &=& 0\\
\label{eq:BC2}\partial^2_s\vec{r}_{\perp,p}(s)\bigr|_{s=\pm1/2} &=& 0.
\end{eqnarray}
\end{subequations}
Here eq. \eqref{eq:BC1} is a consequence of the very definition of the polymer-fixed reference frame, whereas \eqref{eq:BC2} arises from the physical requirement of a vanishing torque at both ends of the polymer. Within this formulation of the problem the WBR approximation holds true whenever $\alpha\ll1$, which may be realized either by virtue of sufficiently strong confinement strengths $c$ or by virtue of sufficiently small chain flexibilities $\epsilon$. The exponential in eq. \eqref{eq:PILT} then takes the form
\begin{equation}
\label{eq:expref}
\begin{split}
&\mathcal{E}[\vec{r}_p(s),\xi,\theta,\vec{u}]=\\&\frac{1}{2\epsilon}\left\{\vec{r}_{\perp,p}''\vec{r}_{\perp,p}'\bigr|_{-\frac{1}{2}}^{\frac{1}{2}}-\vec{r}_{\perp,p}'''\vec{r}_{\perp,p}\bigr|_{-\frac{1}{2}}^{\frac{1}{2}}\right\}+\frac{\xi}{2}\vec{r}_{\perp,p}'\vec{r}_{\perp,p}\bigr|_{-\frac{1}{2}}^{\frac{1}{2}}+\\
&+\frac{1}{2}\Bigl\langle\vec{r}_{p}\Bigl|\mathbb{P}_{\perp}^t\left(\epsilon^{-1}\partial^4_s-\xi\partial^2_s\right)\mathbb{P}_{\perp}+\mathbb{T}^t\frac{c^4}{\epsilon}\mathbb{T}\Bigr|\vec{r}_{p}\Bigr\rangle,
\end{split}
\end{equation}
where $\mathbb{P}_{\perp}\vec{r}_p=\vec{r}_{\perp,p}$ and
\begin{equation}
\label{eq:T}
\mathbb{T}\,\vec{r}_p=\left(
\begin{array}{c c c}
\cos\theta & 0 & \sin\theta\\
0 & 1 & 0
\end{array}
\right)\vec{r}_p+\vec{u}
\end{equation}
gives the transverse displacement vector with respect to the channel-fixed frame $\Sigma_c$. Here $\theta$ is the angle between the $z$-axes of the two reference frames, and $\vec{u}=(u_x,u_y)^t$ connects the origins of $\Sigma_c$ and $\Sigma_p$. Figure \ref{fig:intro} illustrates the transformation in eq. \eqref{eq:T}. An expansion of the transverse polymer fluctuations $\vec{r}_{\perp,p}$ in terms of simple sine-modes
\begin{equation}\label{eq:Expansion}
\vec{r}_{\perp,p}(s)=\sqrt{2}\sum_{n=1}^{\infty}\vec{a}_n\sin\left(n\pi \left(s+\frac{1}{2}\right)\right)
\end{equation}
renders the exponential in eq. \eqref{eq:expref} Gaussian and, in addition, is consistent with the boundary conditions within $\Sigma_p$, as stated in eqs. \eqref{eq:BC}.

Note, that we were able to preserve \textit{hinged ends} boundary conditions and all the convenient features brought about by these at the expense of dealing with two reference frames, rather than a single one. Technically this means, that we have to deal with additional degrees of freedom, embodied by the rotation angle $\theta$ and the displacement vector $\vec{u}$, which would be absent if we stipulated \textit{hinged} or \textit{torqued ends} boundary conditions within $\Sigma_c$. There is a conceptual difference between the latter degrees of freedom and the degrees of freedom represented by the expansion coefficients $\{\vec{a}_n\}$ in eq. \eqref{eq:Expansion}. While variation of the expansion coefficients $\{\vec{a}_n\}$ at constant values of $(\vec{u},\theta)$ leads to deformations of the contour at fixed relative position of the two sets of axes $\Sigma_c$ and $\Sigma_p$, altering the values of $(\vec{u},\theta)$ while fixing $\{\vec{a}_n\}$ amounts to shifting a frozen contour with respect to the channel. Therefore, in what follows we will refer to the coefficients $\{\vec{a}_n\}$ as \textit{``internal modes''}  and we will reserve the term \textit{``global modes''}  for the entities $(\vec{u},\theta)$. In the following, this conceptual difference will help us illuminate the different statistics due to different boundary conditions.

For the time being we concentrate on the \textit{internal modes}, treating the \textit{global modes} as parameters. We thus calculate
\begin{equation}
\label{eq:L}
\mathcal{L}(\xi,\theta,\vec{u})=\frac{1}{\mathcal{Z}}\int\left[\prod_{n=1}^{\infty}d^2\vec{a}_n\right]\exp\left(-\mathcal{E}[\{\vec{a}_n\},\xi,\theta,\vec{u}]\right),
\end{equation}
which is the Laplace transform of the joint probability distribution function $\mathcal{G}(r,\theta,\vec{u})$.
The integrals occurring in eq. \eqref{eq:L} are Gaussian and may readily be performed. It should be noted, however, that the resulting expression is a rather involved function of the \textit{global modes}, which is why we resort to the following expansion scheme in order to render upcoming $\cos\theta$- and $\vec{u}$-integrations amenable to analytical calculations:
\begin{equation}
\label{eq:expansion}
\begin{split}
\mathcal{L}(\xi,\theta,\vec{u})&=\frac{1}{\mathcal{Z}}\;e^{-\frac{c^4}{24\epsilon}\sin^2\theta}e^{-\frac{c^4}{2\epsilon}\vec{u}^2}L^{(0)}(\xi)\times\\
&\times\Bigl\{1+L^{(1)}_{s}(\xi)\,\sin^2\theta+L^{(1)}_{u}(\xi)\,\vec{u}^2+\dots\Bigr\}.
\end{split}
\end{equation}
Here we have chosen to perform an expansion with respect to the small quantities $\sin^2\theta$ and $\vec{u}^2$ in order to account for the symmetry properties of the system under study. Also note that the expansion in eq. \eqref{eq:expansion} explicitly keeps track of the rigid rod limit $\epsilon\rightarrow0$, where $\mathcal{L}\propto e^{-\frac{c^4}{24\epsilon}\sin^2\theta}e^{-\frac{c^4}{2\epsilon}\vec{u}^2}$. In what follows we will calculate the effect of the \textit{global modes} to leading order. 

\begin{figure}
	\centering
		\includegraphics[width=.45\textwidth]{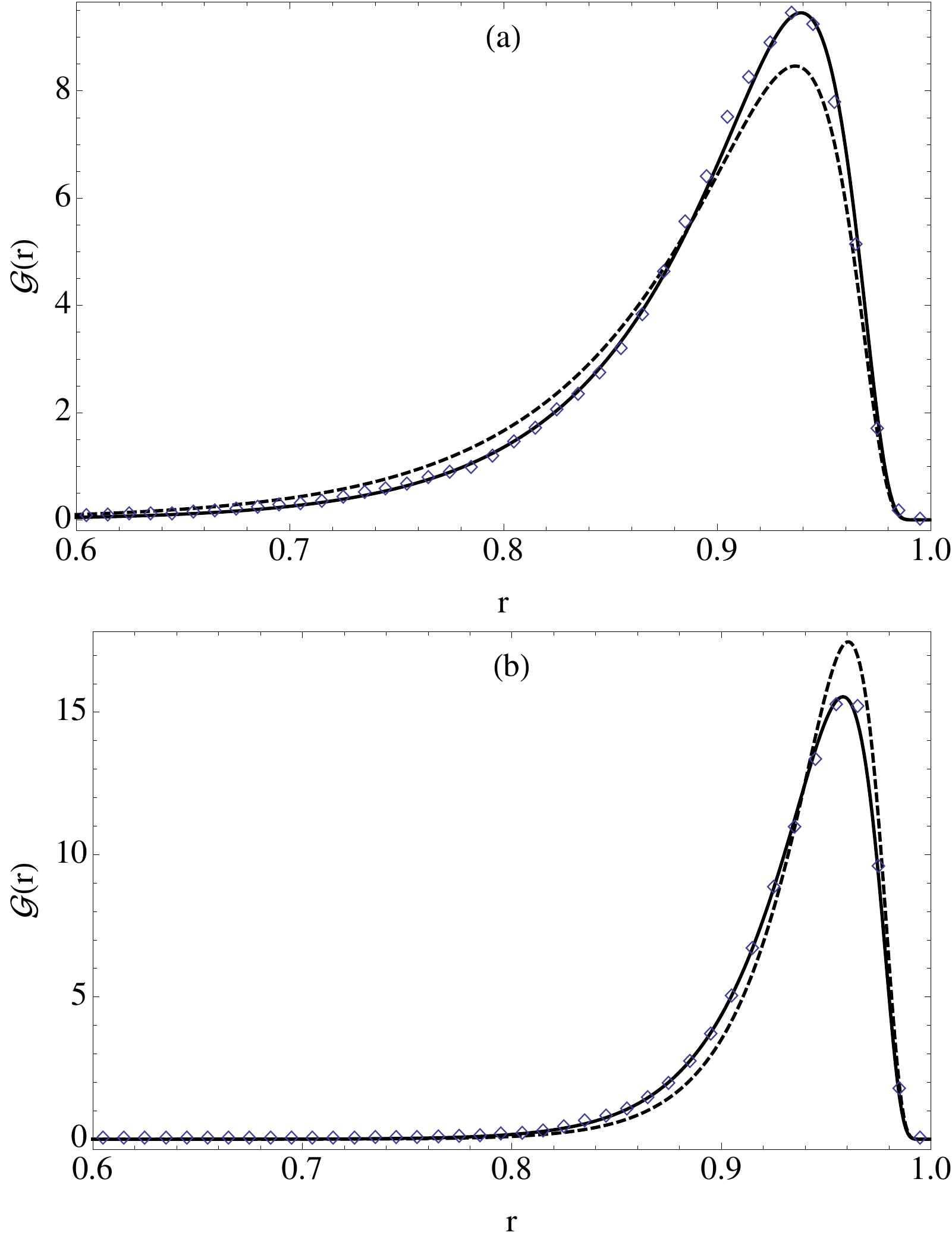}
	\caption{Comparison of $\mathcal{G}(r)$ (eq. \eqref{eq:RDF}, solid lines) and $\mathcal{G}_h(r)$ (eq. \eqref{eq:LeviMecke}, dashed lines) with MC simulation data (data points) for weak and intermediate confinement strengths.
(a) Weak confinement ($c=0.5$, $\epsilon=0.7$). For entropic reasons the \textit{global modes} shift the end-to-end distance statistics towards larger values of $r$.
(b) Intermediate confinement ($c=3.0$, $\epsilon=0.5$). The energy optimizing effect of the \textit{global modes} shifts the end-to-end distance statistics towards lower values of $r$.}
	\label{fig:RDF}
\end{figure}

Using eqs. \eqref{eq:expref} and \eqref{eq:L}, the zeroth and first order coefficients in expansion \eqref{eq:expansion} are found to be
\begin{subequations}
\label{eq:coefficients}
\begin{eqnarray}
L^{(0)}(\xi)                  &=& \prod_{p=1}^{\infty}\frac{\xi_p}{\xi_p-\xi}\\
L^{(1)}_{s}(\xi) &=& \frac{c^4}{2\epsilon}\sum_{p=1}^{\infty}\frac{1}{(p\pi)^2(\xi-\xi_p)}+\nonumber\\
                 &&  +\frac{c^8}{\epsilon^2}\sum_{2\mathbb{N}}\frac{1}{(p\pi)^4(\xi-\xi_p)}\\
L^{(1)}_{u}(\xi)         &=& \frac{4c^8}{\epsilon^2}\sum_{2\mathbb{N}+1}\frac{1}{(p\pi)^4(\xi-\xi_p)},
\end{eqnarray}
\end{subequations}
where we abbreviated
\begin{equation}
\label{eq:Fn}
\xi_p\equiv-\frac{(p\pi)^2}{\epsilon}-\frac{c^4}{(p\pi)^2\epsilon}.
\end{equation}
At this point, it is worth noting that $\mathcal{L}(\xi,\theta,\vec{u})$ is intimately related to the moment generating function $M(\xi)\equiv\langle e^{\xi r}\rangle\equiv\int dr\mathcal{G}(r)e^{\xi r}$. Using $\mathcal{L}(\xi,\theta,\vec{u})=\mathfrak{L}\left[\mathcal{G}(r,\theta,\vec{u})\right]$ and\footnote{Here and in the following the constant factor $2\pi$, arising from the (trivial) azimuthal integration, is understood to be integrated into the normalization constant.}
\begin{equation}
\label{eq:GG}
\mathcal{G}(r)=r^2\int d(\cos\theta)\int d^2\vec{u}\,\mathcal{G}(r,\theta,\vec{u}) 
\end{equation}
it is a straightforward matter to show that
\begin{equation}
\label{eq:ML}
M(\xi)=\partial^2_{\xi}\left[e^{\xi}\mathcal{L}(\xi)\right],
\end{equation}
where $\mathcal{L}(\xi)\equiv\int d(\cos\theta)\int d^2\vec{u}\mathcal{L}(\xi,\theta,\vec{u})$.
Hence, knowledge of $\mathcal{L}(\xi)$ directly provides us with the $n^{\text{th}}$ order moments of the end-to-end distance $r$ without explicit recourse to the RDF:
\begin{equation}
\label{eq:moments}
\langle r^n\rangle=\partial^{n+2}_{\xi}\left[e^{\xi}\mathcal{L}(\xi)\right]\Bigr|_{\xi=0}.
\end{equation}
The RDF itself follows from $\mathcal{L}(\xi)$ by means of an inverse Laplace transform. Using eq. \eqref{eq:GG} and the definition of $\mathcal{L}(\xi)$ we find
\begin{equation}
\label{eq:G}
\mathcal{G}(r)=r^2\mathfrak{L}^{-1}\left[\mathcal{L}(\xi)\right],
\end{equation}
where we used the linearity of the inverse Laplace transform. Inspection of eqs. \eqref{eq:expansion} and \eqref{eq:coefficients} shows that $\mathcal{L}(\xi)$ contains poles of second order at $\{\xi_p\}_{p\in\mathbb{N}}$. Thence we may invoke residues calculus in order to perform the inverse Laplace transform \eqref{eq:G} and write down $\mathcal{G}(r)$ as a sum over poles. Invoking our hitherto derived results we find to the desired order in the \textit{global modes}
\begin{equation}
\label{eq:RDF}
\mathcal{G}(r)=\frac{r^2}{\mathcal{Z}}\sum_{p=1}^{\infty}\mathcal{G}^{(0)}_p(r)\,\left\{1+\frac{\epsilon}{c^4}\left[A\,\mathcal{G}^{(1)}_{s,p}+2\,\mathcal{G}^{(1)}_{u,p}\right]\right\}.
\end{equation}
Here the following terms have been defined:
\begin{subequations}
\label{eq:Gs}
\begin{eqnarray}
\label{eq:G0}
\mathcal{G}^{(0)}_p(r)   &=&   -\xi_pe^{\xi_p\rho}\prod_{n\neq p}\frac{\xi_n}{\xi_n-\xi_p}\\
\mathcal{G}^{(1)}_{s,p}(r) &=& \frac{c^4}{\epsilon(p\pi)^2}\left(\rho-S^{(0)}_{\mathbb{N},p}\right)\left(\frac{1}{2}+\frac{c^4}{\epsilon(p\pi)^2}\delta_{p,2\mathbb{N}}\right)+\nonumber\\
&&+\frac{c^4}{2\epsilon}S^{(2)}_{\mathbb{N},p}+\frac{c^8}{\epsilon^2}S^{(4)}_{2\mathbb{N},p}\\
\mathcal{G}^{(1)}_{u,p}(r) &=& 4\frac{c^8}{\epsilon^2}\left[\frac{\delta_{p,2\mathbb{N}+1}}{(p\pi)^4}\left(\rho-S^{(0)}_{\mathbb{N},p}\right)+S^{(4)}_{2\mathbb{N}+1,p}\right]
\end{eqnarray}
\end{subequations}
and
\begin{equation}
\label{eq:A}
A = 12\left(1-\frac{x}{D(x)}+2x^2\right)\Biggr|_{x=\sqrt{c^4/(24\epsilon)}}.
\end{equation}
In eqs. \eqref{eq:Gs} the sums
\begin{equation}
S^{(k)}_{\mathbb{A},p}\equiv\sum_{n\in \mathbb{A},n\neq p}\frac{1}{(n\pi)^k(\xi_p-\xi_n)}
\end{equation}
may actually be represented in terms of transcendental functions by means of an expansion in partial fractions (not shown). The function $D(x)$, occurring in eq. \eqref{eq:A} denotes Dawson's integral, $D(x)\equiv e^{-x^2}\int_0^xdy\,e^{y^2}$.

As an aside we note that due to the fact that the exponential in eq. \eqref{eq:G0} has a local maximum at $q_p\equiv p\pi= c$ ($c>0$), wavenumbers $q_p$ in the vicinity of the inverse Odijk length $L_d$ contribute the most to the sum in eq. \eqref{eq:RDF}. This is, of course, consistent with the observation that the polymer preferably stores length at wavenumbers $q\approx c=L_d^{-1}$.

\begin{figure}
	\centering
		\includegraphics[width=.45\textwidth]{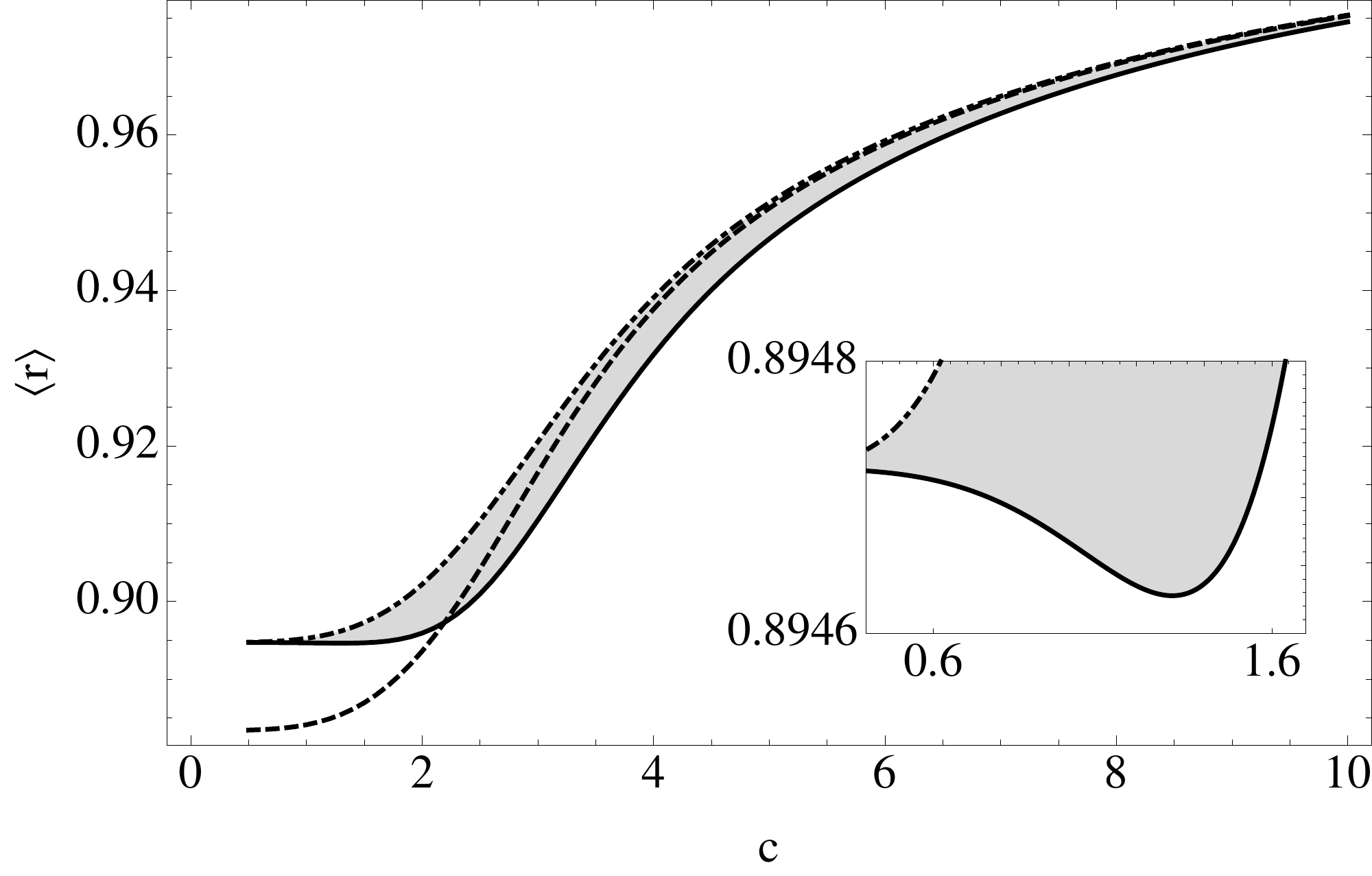}
	\caption{Mean end-to-end distance $\langle r\rangle$ as a function of confinement strength $c$, calculated on the basis of $\mathcal{G}(r)$ (eq. \eqref{eq:RDF}, solid line), $\mathcal{G}_h(r)$ (eq. \eqref{eq:LeviMecke}, dashed line) and $r^2\mathcal{G}_h(r)$ (dot-dashed line). The shaded area gives an illustrative measure of the impact of the energy optimizing effect. Inset: Only the first order corrected model exhibits a local minimum of $\langle r\rangle(c)$ at onsetting confinement.}
	\label{fig:rmean}
\end{figure}

Repeating our calculations, stipulating rigid coincidence of polymer-fixed and channel-fixed reference frames---\textit{\textit{i.e.}} invoking \textit{hinged ends} boundary conditions within $\Sigma_c$---we are left with the following simplified version of the RDF
\begin{equation}\label{eq:LeviMecke}
\mathcal{G}_h(r)=\mathcal{Z}_h^{-1}\,\sum_{p=1}^{\infty}\mathcal{G}^{(0)}_p(r),
\end{equation}
which was determined before by Levi and Mecke in ref. \cite{levi-2007-78}. Comparing eqs. \eqref{eq:RDF} and \eqref{eq:LeviMecke}, suggests to think of the RDF stated in eq. \eqref{eq:RDF} as being made up of three structural parts: first, the coefficients $\mathcal{G}^{(0)}_p(r)$, carrying all the statistical information related to the \textit{internal modes}. Second, the common factor $r^2$, and, third, the terms in curly brackets, both arising as a consequence of the \textit{global modes}. Note that the global factor $r^2$ is simply a phase space factor and a consequence of the spherical integrations performed, whereas the non-trivial correction factors in curly brackets are brought about by the explicit $(\theta,\vec{u})$-dependence of the confinement potential $\mathcal{H}_{\text{pot}}$.

Depending on the confinement strength, the three structural parts have different impact on the chain's end-to-end distance statistics. Figure \ref{fig:RDF}, comparing eqs. \eqref{eq:RDF} and \eqref{eq:LeviMecke}, illustrates this point.

In the regime of vanishing (and weak) confinement, the \textit{global modes}\footnote{more precisely the quantities $\cos\theta$ and $\vec{u}$} are uniformly distributed, whence the non-trivial correction terms $\mathcal{G}^{(1)}_{s/u,p}$ in eq. \eqref{eq:RDF} vanish and $\mathcal{G}(r)$ reduces to the free-space solution determined in ref. \cite{wilhelm-1996-77}. In this context note that our expansion scheme reproduces the zero-confinement limit (where the expansion parameters become particularly large) even by zeroth order. Hence, our first order corrected model is supposed to give an excellent approximation for virtually any choice of the confinement parameter $c$, safe for a very narrow regime of onsetting confinement, where the expansion parameters are still large and where the non-trivial effects of confinement are already perceptible. We will return to this point below. 
For now the important point to emphasize is that in the limit of vanishing confinement the only fingerprint left behind by the \textit{global modes} is embodied by the common phase space factor $r^2$, which is lost by any approach where the positions and/or orientations of the polymer's ends are artificially constrained. From a physical point of view, the phase space factor $r^2$ arises for entropic reasons, since in 3 dimensional embedding space the number of states consistent with a given contour $\{a_n\}$, with end-to-end distance $r$, is proportional to the surface of a sphere with the same radius, and therefore to $r^2$. Neglecting this factor enhances the probability of comparatively short configurations and thus shifts the mean end-to-end distance $\langle r\rangle$ below its actual value, which is exactly what can be observed in figs. \ref{fig:RDF}(a) and \ref{fig:rmean}. Apart from the simple factor $r^2$, however, eqs. \eqref{eq:RDF} and \eqref{eq:LeviMecke} are identical in the limit $c\rightarrow0$, so that the shortcomings of $\mathcal{G}_h(r)$ may easily be corrected.

In the regime of intermediate confinement (fig. \ref{fig:RDF}(b)), eqs. \eqref{eq:RDF} and \eqref{eq:LeviMecke} still deviate from each other, but this time the 
simple factor $r^2$ alone is no longer capable of refining the behavior of $\mathcal{G}_h(r)$. Here, the essential corrections to be taken into account are contained in the nontrivial terms $\mathcal{G}^{(1)}_{s/u,p}$. In this case, the presence of the confining channel walls noticeably constrains the thermal undulations of the polymer. For any ``frozen'' contour, \textit{\textit{i.e.}} for any fixed set $\{\vec{a}_n\}$, however, the \textit{global modes} allow for a shift of the whole contour with respect to the channel walls in such a way as to lower the potential energy of the entire system. This energy-optimizing effect, brought about by the \textit{global modes} and analytically contained within the nontrivial coefficients $\mathcal{G}^{(1)}_{s/u,p}$, enhances the probability of relatively compact conformations. Neglecting these coefficients inevitably leads to an overestimation of the stretching effect due to the confining walls, which is precisely what can be observed in figs. \ref{fig:RDF}(b) and \ref{fig:rmean}. Moreover, note that since this effect mainly results from the topological restrictions imposed by the channel, the difference in chain statistics arising from the different conditions at the boundaries of the polymer is nearly independent of the chain's flexibility in this regime. 

Finally, once confinement is sufficiently strong, the \textit{global modes} become so effectively constricted, that they finally lose their significance for the determination of the RDF, and the difference between $\mathcal{G}(r)$ and $\mathcal{G}_h(r)$ vanishes (not shown). Both functions approach the same limit, which is a delta-function at full extension.

In addition to the quantitative differences discussed so far, the particular choice of boundary conditions also impacts the qualitative behavior of $\langle r\rangle$ as a function of the confinement strength. Figure \ref{fig:rmean} illustrates this point. Whereas hinging the polymer's ends to the channel axis implies a monotonically increasing end-to-end distance, accounting for the genuine conditions at the boundaries of the chain gives rise to a local minimum in $\langle r\rangle(c)$ at onsetting confinement. Although our first order corrected model is not capable of reproducing this feature in quantitative detail, our approach unambiguously traces back this phenomenon---which has recently been reported in MC simulations \cite{wagner-2007-75,cifra}---to the influence of the \textit{global modes}.

In conclusion, we gave an explicit, analytical account for the derivation of the radial distribution function (RDF), as well as for the $n^{\text{th}}$ order moments $\langle r^n\rangle$ of the end-to-end distance, for semiflexible polymers in cylindrical confinements. Here we particularly focused on the use of proper boundary conditions and highlighted their influence within different confinement scenarios.
To this end,  we introduced additional \textit{global modes} in order to account for the possibility of the chain to perform global movements within the confining channel, and calculated their influence on the polymer's radial statistics to leading order. In a regime of weak and vanishing confinement the \textit{global modes} are trivially distributed and their effect on the RDF reduces to a simple phase space factor, which is of purely entropic origin. In contrast, in a regime of intermediate confinement the system's potential energy explicitly depends on the \textit{global modes}, which therefore affect the RDF in a non-trivial way. We found that neglecting their influence in the latter regime inevitably results in an overestimation of the stretching effect due to confinement. Finally, in addition to these quantitative differences the \textit{global modes} were found to be responsible for the formation of a local minimum in $\langle r\rangle(c)$ at onsetting confinement.

\acknowledgments
It is a great pleasure to acknowledge helpful discussions and support by Benedikt Obermayer and Wolfram M\"obius, as well as financial support by the Deutsche Forschungsgemeinschaft Contract No. FR 850/8--1.

\end{document}